\documentclass[12pt] {article}
\usepackage{latexsym}
\usepackage{amsmath}
\usepackage{mathrsfs}
\usepackage{amsfonts}
\usepackage{graphicx}

\usepackage[totalheight = 21cm, totalwidth = 17cm]{geometry}
\newcommand{\bea}{\begin{eqnarray}}
\newcommand{\eea}{\end{eqnarray}}
\newcommand{\be}{\begin{equation}}
\newcommand{\ee}{\end{equation}}

\begin{document}
\begin{titlepage}  
\baselineskip=21pt
\vspace{2cm}
\rightline{\tt hep-th/0412240}
\rightline{CERN-PH-TH/2004-264}
\rightline{MIFP-04-27, ACT-06-04}
\vspace{1cm}
\begin{center}
{\bf {\large Cosmic Acceleration and the String Coupling }}
\end{center}
\begin{center}
\vskip 0.2in
{\bf John~Ellis}$^1$, {\bf N. E. Mavromatos}$^{2}$ and 
{\bf D. V. Nanopoulos}$^{3}$
\vskip 0.1in
{\it
$^1${TH Division, Physics Department, CERN, CH-1211 Geneva 23, Switzerland}\\
$^2${Theoretical Physics, Physics Department, 
King's College London, Strand WC2R 2LS, UK}\\
$^3${George P. and Cynthia W. Mitchell Institute for Fundamental 
Physics, \\
Texas A\&M University, College Station, TX 77843, USA; \\
Astroparticle Physics Group, Houston
Advanced Research Center (HARC),
Mitchell Campus,
Woodlands, TX~77381, USA; \\
Academy of Athens,
Division of Natural Sciences, 28~Panepistimiou Avenue, Athens 10679,
Greece}\\
} 

\vspace{0.5cm}
{\bf Abstract}
\end{center}
\baselineskip=18pt \noindent  

In the context of a cosmological string model describing the propagation
of strings in a time-dependent Robertson-Walker background space-time, we
show that the asymptotic acceleration of the Universe can be identified
with the square of the string coupling. This allows for a direct
measurement of the ten-dimensional string coupling using cosmological
data.  We conjecture that this is a generic feature of a class of
non-critical string models that approach asymptotically a conformal
(critical) $\sigma$ model whose target space is a four-dimensional
space-time with a dilaton background that is linear in $\sigma$-model
time. The relation between the cosmic acceleration and the string coupling
does not apply in critical strings with constant dilaton fields in four
dimensions.

\vfill
\leftline{CERN-PH-TH/2004-264}
\leftline{December 2004}
\end{titlepage}
\baselineskip=18pt

String theory~\cite{gsw,polchinski} was first developed as a theory of the
strong interactions, but it soon turned out that mathematical consistency
(world-sheet conformal invariance) required the theory to live in
higher-dimensional space times. Even target-space supersymmetry was not
successful in reducing the number of space-time dimensions below ten.
Thus, enormous effort has been expended on the compactification of the
extra dimensions, with the eventual aim of accommodating the Standard
Model at low energies. Many ways were found to construct low-energy models
that could be consistent with the current particle physics phenomenology,
but string models of this type had zero predictability, in the sense that
they were unable to make predictions for the parameters of the Standard
Model, and there were many string models with indistinguishable low-energy
limits.

Although in principle string theory has no free parameters, and the ground
state corresponding to the observable low-energy world is supposed to be
chosen dynamically, a detailed understanding of mechanism for choosing the
ground state has not been achieved so far. Lacking a microscopic,
dynamical mechanism for specifying the various string model parameters,
such as the compactification radii and the four-dimensional gauge
couplings, one has had simply to fix them by hand, so as to match the
results with experimental observations in particle physics. In this
framework, the mechanism whereby one particular model is chosen from among
the complicated string `landscape'~\cite{Land} is still unclear.

The most important and fundamental string parameter is the string
coupling, $g_s$, which determines the regime of validity of string
perturbation theory, and hence the world-sheet $\sigma$-model scheme for
low-energy computations of the low-energy string effective action. Since
$g_s$ is connected to the unified ten-dimensional gauge coupling of the
effective supersymmetric low-energy theory, its value is usually inferred
from particle phenomenology.
The string coupling is not a constant but, like any other dynamical 
coupling in a supersymmetric field theory,
is related to the vacuum expectation value of a field,
in this particular case the dilaton field $\Phi$, which belongs to 
the gravitational multiplet obtained from the string~\cite{gsw}
\begin{equation}
g_s^2 = e^{2\Phi}.
\label{defstringcoupl}
\end{equation}
Usually, upon compactification the dilaton field is split
into a product of two factors, one depending on the compact six-dimensional 
space coordinates and the other on the four-dimensional space-time 
coordinates, which are supposed to correspond to 
the large, uncompactified coordinates
of our observable world. In most of the phenomenological approaches 
to model building, the four-dimensional 
dilaton field has been assumed to be constant and therefore trivial, since 
this constant value could be absorbed in an innocuous shift in the field. 

In this approach, neither the string coupling nor the unified gauge
coupling are accessible directly to experimental measurement. It is
consistency of the available phenomenological model with low-energy
observational data that leads to an indirect fixing of the string
coupling. A popular value is $g_s^2 \simeq 0.52$, which, upon
compactification to small dimensions (of the order of a tenth of the
four-dimensional Planck mass, $M_P \sim 10^{19}$ GeV), yields a
four-dimensional unified gauge coupling strength $g_U^2/4\pi \sim 1/24$ at
scales $M_U \sim 10^{16}$ GeV, as suggested by extrapolating the measured
gauge couplings to high energies in the context of the minimal
supersymmetric extension of the Standard Model.
  
Modern developments in string theory~\cite{polchinski} make possible
consistent quantum treatments of domain-wall structures in string theory
(D-branes). These have opened up novel ways of looking at both the
microcosmos and the macrocosmos, offering new insights into both particle
phenomenology and the cosmic evolution of our Universe. In the
microcosmos, there are novel ways of compactification, either via the
observation~\cite{add} that large (compared to the string scale)  extra
dimensions are consistent both with the foundations of string theory and
phenomenology, or by viewing our four-dimensional world as a brane
embedded in a bulk space-time. This would allow for large extra bulk
dimensions, which could even be infinite in size~\cite{randal}, offering
new ways to analyze the large hierarchy between the Planck scale and the
electroweak symmetry-breaking scale. In this modern approach, fields in
the gravitational (super)multiplet of the (super)string are allowed to
propagate in the bulk, but not the gauge fields, which are attached to the
brane world. In this way, the weakness of gravity as compared to the
rest of the interactions is a result of the large extra dimensions. Their
compactification is not necessarily achieved through conventional
means, i.e., closing up the extra dimensions in compact spatial manifolds,
but might also involve shadow brane worlds with special reflecting
properties (such as orientifolds), which bound the bulk
dimension~\cite{ibanez}. In such approaches, the string scale $M_s$ is not
necessarily identical to the four-dimensional Planck mass scale $M_P$, but
instead they are related through the large compactification volume $V_6$:
\begin{equation}
M_P^2 = \frac{8M_s^8 V_6}{g_s^2}.
\label{planckstring}
\end{equation}
As for the macrocosmos, there are novel ways of discussing cosmology
in brane worlds, which may revolutionize our way of approaching
issues such as inflation~\cite{langlois,ekpyrotic}. 

Mounting experimental evidence from diverse astrophysical sources presents
important cosmological puzzles that string theory must address if it is to
provide a realistic description of Nature. Observations of large-scale
structures, distant Type-1asupernovae~\cite{snIa}, and the cosmic
microwave background fluctuations (by WMAP~\cite{wmap} in particular) have
established that the Hubble expansion of our Universe is currently
accelerating, and that 70\% of its energy density consists of unknown dark
energy that appears in `empty' space and does not clump with ordinary
matter.

These observations have great potential significance for string theory,
and may even revolutionize the approach to it that has normally been
followed so far. If the dark energy leads to an asymptotic de Sitter
horizon, as would occur if it turns out to be a true cosmological
constant, then the entire concept of the scattering S-matrix breaks down,
and hence the conventional approach to string theory. On the other hand,
if there is some quintessential mechanism for relaxing the vacuum energy,
so that the vacuum energy density vanishes at large cosmic times in a
manner consistent with the existence of an S-matrix, there is still the
open issue of embedding such models in (perturbative) string theory. One
would need, in particular, to develop a consistent $\sigma$-model
formulation of strings propagating in such time-dependent, relaxing
space-time backgrounds.

{\it We here propose a resolution of this dilemma, based on string theory
in a time-dependent dilaton background, in which the asymptotic
acceleration of the Universe is directly related to the string coupling.}

The world-sheet conformal-invariance conditions of critical string theory 
are equivalent to the target-space equations of motion for the
background fields through which the string propagates. These conditions 
are very 
restrictive,
allowing only for vacuum solutions of (critical) strings to be constructed
in this way. The main problem may be expressed as follows. 
Consider the graviton world-sheet $\beta$ function, which is nothing but 
the Ricci tensor of the target space-time background to lowest order 
in $\alpha '$:
\begin{equation} 
\beta_{\mu\nu} = \alpha ' R_{\mu\nu}~,
\label{beta} \end{equation}
where we ignore the possible presence of other fields, for simplicity.
Conformal invariance requires the vanishing condition $\beta_{\mu\nu}=0$,
which implies that the background must be Ricci flat, which is a solution
of the vacuum Einstein equations.
The issue then arises how to describe in string theory cosmological
backgrounds, which are not vacuum solutions, but require the presence of a
matter fluid and hence a non-vanishing Ricci tensor.  In this respect, we
see that a cosmological constant is inconsistent with the conformal
invariance of string since, for instance, a de Sitter Universe with a
positive cosmological constant $\Lambda > 0$ has a non-zero Ricci tensor
$R_{\mu\nu} = \Lambda g_{\mu\nu}$, where $g_{\mu\nu}$ is the metric
tensor.

An interesting proposal for obtaining a non-zero cosmological constant in
string theory was made in~\cite{fischler}. It was suggested that dilaton
tadpoles in higher-genus world-sheet surfaces, which produce additional
modular infinities whose regularization leads to extra world-sheet
structures in the $\sigma$-model not appearing at the world-sheet level,
modify the string $\beta$-function in such a way that the Ricci tensor of
the space-time background is now that of a de Sitter Universe, with a
cosmological constant specified by the dilaton tadpole graph. The problem
with this approach is the above-mentioned existence of an asymptotic
horizon in the de Sitter case, which prevents the proper definition of
asymptotic states, and hence an S-matrix.  Since the perturbative
world-sheet formalism is based on the existence of such an S-matrix, there
is a question of consistency in this approach.

It was proposed in~\cite{aben} that a way out of this difficulty would be 
to assume specific time-dependent dilaton backgrounds, with a linear
dependence on time in the so-called $\sigma$-model frame:
\begin{equation}
\label{lineardil}
\Phi = {\rm const} - Q~t
\end{equation}
where $Q$ is a constant, and $Q^2 > 0$ is a deficit in the 
$\sigma$-model central charge. Such
backgrounds, even when the $\sigma$-model metric is flat, lead to exact
solutions (in all orders in $\alpha '$) of the conformal-invariance
conditions of the pertinent stringy $\sigma$-model, thereby constituting
acceptable solutions from a perturbative string viewpoint. 
The appearance of $Q$ allowed this 
supercritical string theory~\cite{aben}
to be formulated in spaces with numbers of dimensions different from the 
critical case.
This was actually the first example of a non-critical string, with the 
target-space coordinates $X^i$, $i=1, \dots D-1$, 
playing the r\^oles of the pertinent $\sigma$-model fields.
This non-critical string was not conformally invariant, and hence 
required Liouville 
dressing~\cite{ddk}. The Liouville field had time-like signature
in target space, since the central charge deficit $Q^2$ was positive
in the model of \cite{aben}, and its zero mode played the r\^ole of target 
time.

As a result of the existence of a non-trivial dilaton field, the 
Einstein term in the effective $D$-dimensional low-energy
field theory action is conformally rescaled by $e^{-2\Phi}$.
This requires a specific redefinition of target time in order that the
metric acquires the standard Robertson-Walker (RW) form in the normalized
Einstein frame for the effective action:
\begin{equation}
ds^2_E = -dt_E^2 + a_E(t_E)^2 \left(dr^2 + r^2 d\Omega^2 \right),
\end{equation}
where we have only exhibited a spatially-flat RW metric for 
definiteness,
and $a_E(t_E)$ is an appropriate scale factor, which is simply a function 
of 
the Einstein-frame time $t_E$
in the homogeneous cosmological backgrounds that we assume throughout.

The Einstein-frame time is related to the $\sigma$-model-frame 
time~\cite{aben} by:
\begin{equation}
\label{einsttime}
dt_E = e^{-\Phi}dt \qquad \to \qquad t_E = \int ^t e^{-\Phi(t)} dt~. 
\end{equation} 
The linear dilaton background (\ref{lineardil}) yields then
the following relation between the Einstein- and $\sigma$-model-frame
times:
\begin{equation} 
t_E = c_1 + \frac{c_0}{Q}e^{Qt},
\end{equation}
where $c_{1,0}$ are appropriate (positive) constants.
Thus, a dilaton background that is
linear in $\sigma$-model-frame time (\ref{lineardil}) 
will scale logarithmically with 
the Einstein-frame time $t_E$, which is just the Robertson-Walker cosmic 
time:
\begin{equation}
\label{dil2}
\Phi (t_E) = {\rm const.'} - {\rm ln}(\frac{Q}{c_0}t_E).
\end{equation} 
In this regime, the string coupling (\ref{defstringcoupl}) 
varies with the cosmic time $t_E$ as:
\begin{equation}
g_s^2 (t_E) \propto \frac{1}{t_E^2},
\end{equation}
implying that the effective string coupling 
vanishes asymptotically in cosmic time. 
In the linear-dilaton background of~\cite{aben}, the asymptotic
space-time metric in the Einstein frame reads:
\begin{equation} 
\label{metricaben}
ds^2 = -dt_E^2 + a_0^2 t_E^2 \left(dr^2 + r^2 d\Omega^2 \right),
\end{equation}
where $a_0$ a constant, which is a linearly-expanding Universe. Clearly, 
there is no acceleration in the Universe (\ref{metricaben}).

In~\cite{emn} we went one step further than the analysis in~\cite{aben},
and considered more complicated $\sigma$-model metric backgrounds, which
did not satisfy the $\sigma$-model conformal-invariance conditions, and
therefore were in need of Liouville dressing in order to restore conformal
invariance. Such backgrounds were also allowed to be time-dependent, and
the target time was identified with the Liouville world-sheet zero mode,
thereby not increasing the target space-time dimensionality. We have
provided several justifications and checks of this
identification~\cite{emn}, which is possible only when the initial
$\sigma$-model is supercritical, so that the Liouville mode has time-like
signature~\cite{aben,ddk}. For example, in certain
models~\cite{gravanis,brany}, such an identification was energetically
preferable from a target-space viewpoint, since it minimized certain
effective potentials in the low-energy field theory corresponding to the
string theory at hand.

Such non-critical $\sigma$ models relax asymptotically in cosmic Liouville
time to conformal $\sigma$ models, the latter viewed as equilibrium points
in string theory space. In some interesting cases of relevance to
cosmology, which were particularly generic, the asymptotic conformal field
theory was that of~\cite{aben}, with a linear dilaton and a flat Minkowski
target-space metric in the $\sigma$-model frame.

One such model was considered in detail in~\cite{dgmpp}. The model was
originally formulated within a specific string theory, namely
ten-dimensional Type-0~\cite{type0}, which leads to a non-supersymmetric
target-space spectrum, as a result of a special projection of the
supersymmetric partners out of the spectrum. However, the basic properties
of its cosmology, which are those interest to us in in this work, are
sufficiently generic that they can be extended to the bosonic sector of 
any other
effective low-energy supersymmetric field theory of supersymmetric
strings, including those relevant to unified particle physics
phenomenology.

The ten-dimensional metric configuration considered in~\cite{dgmpp} 
was: 
\begin{equation}
G_{MN}=\left(\begin{array}{ccc}g^{(4)}_{\mu\nu} \qquad 0 \qquad 0 \\
0 \qquad e^{2\sigma_1} \qquad 0 \\ 0 \qquad 0 \qquad
e^{2\sigma_2} I_{5\times 5} \end{array}\right),
\label{metriccomp}
\end{equation}
where lower-case Greek indices are four-dimensional space-time
indices, and $I_{5\times 5}$ denotes the $5\times 5$ unit matrix.
We have chosen two different scales for the internal space. The field
$\sigma_{1}$ sets the scale of the fifth dimension, while
$\sigma_{2}$ parametrizes a flat five-dimensional space. In the
context of the cosmological models we treat here, the
fields $g_{\mu\nu}^{(4)}$, $\sigma_{i},~i=1,2$ are assumed to
depend on the time $t$ only.

Type-0 string theory, as well as its supersymmetric versions appearing in
other scenarios including brane models, contains appropriate form fields
with non-trivial gauge fluxes (flux-form fields), which live in the
higher-dimensional bulk space. In the specific model of~\cite{type0}, one
such field was considered to be non-trivial. As was demonstrated
in~\cite{dgmpp}, a consistent background choice for the flux-form field
has the flux parallel to to the fifth dimension $\sigma_1$. This implies
that the internal space is crystallized (stabilized)  in such a way that
this dimension is much larger than the remaining five dimensions
$\sigma_2$, demonstrating the physical importance of the flux fields for
large radii of compactification.

Considering the fields to be time-dependent only, i.e., considering
spherically-symmetric homogeneous backgrounds, restricting ourselves to
the compactification (\ref{metriccomp}), and assuming a Robertson-Walker
form of the four-dimensional metric with scale factor $a(t)$, the
generalized conformal-invariance conditions and the Curci-Pafutti
$\sigma$-model renormalizability constraint~\cite{curci} yield a set of
differential equations which were solved numerically in~\cite{dgmpp}.
The generic form of these equations reads~\cite{ddk,emn,dgmpp}:
\begin{equation} 
  {\ddot g}^i + Q(t){\dot g}^i = -{\tilde \beta}^i,
\label{liouvilleeq}
\end{equation} 
where the ${\tilde \beta}^i$ are the Weyl-anomaly coefficients of the 
stringy $\sigma$-model on the background $\{ g^i \}$. 
In the model of~\cite{dgmpp}, the $\{ g^i \}$ include
graviton, dilaton, tachyon, flux and moduli fields $\sigma_{1,2}$,
whose vacuum expectation values control the sizes of the extra dimensions.

The detailed analysis of~\cite{dgmpp} indicated that the moduli fields
$\sigma_i$ froze quickly to their equilibrium values. Thus, together with
the tachyon field which also decays to a constant value rapidly, they
decouple from the four-dimensional fields at very early stages in the
evolution of this string Universe~\footnote{The presence of the tachyonic
instability in the spectrum is due to the fact that in Type-0 strings
there is no target-space supersymmetry, by construction.  From a
cosmological viewpoint the tachyon fields are not necessarily bad
features, since they may provide the initial instability leading to cosmic
expansion~\cite{dgmpp}.}. There is an inflationary phase in this scenario
and a dynamical exit from it. The important point to guarantee the exit is
the fact that the central-charge deficit $Q^2$ is a time-dependent entity
in this approach, obeying specific relaxation laws determined by the
underlying conformal field theory~\cite{dgmpp,gravanis,brany}. In fact,
the central charge runs with the local world-sheet renormalization group
scale, namely the zero mode of the Liouville field, which is
identified~\cite{emn} with the target time in the $\sigma$-model frame.  
The supercriticality~\cite{aben} $Q^2 > 0$ of the underlying $\sigma$
model is crucial, as already mentioned.  Physically, the non-critical
string provides a framework for non-equilibrium dynamics, which may be the
result of some catastrophic cosmic event, such as a collision of two brane
worlds~\cite{ekpyrotic,gravanis,brany}, or an initial quantum fluctuation.

In the generic class of non-critical string models considered in this
work, the $\sigma$ model always asymptotes, for long enough cosmic times,
to the linear-dilaton conformal $\sigma$-model field theory
of~\cite{aben}. But it is important to stress that this is only an
asymptotic limit. In this respect, the current era of our Universe is
viewed as being close, but still not quite at the relaxation (equilibrium)
point, in the sense that the dilaton is almost linear in the
$\sigma$-model time, and hence varies logarithmically with the
Einstein-frame time (\ref{dil2}). It is expected that this slight
non-equilibrium will lead to a time-dependence of the unified gauge
coupling and other constants such as the four-dimensional Planck length
(\ref{planckstring}) that characterize the low-energy effective field
theory, mainly through the time-dependence of the string coupling
(\ref{defstringcoupl}) that results from the time-dependent linear dilaton
(\ref{lineardil}).

The asymptotic regime of the Type-0
cosmological string model of~\cite{dgmpp} has been obtained 
analytically, by solving the pertinent equations (\ref{liouvilleeq})
for the various fields. As already mentioned, at late times
the theory becomes four-dimensional, and the only non-trivial
information is contained in the scale factor and the dilaton, 
given that the topological flux field remains conformal in this approach,
and the moduli and initial tachyon fields decouple very fast during the 
initial stages after inflation in this model.
For times that are long after the initial fluctuations, 
such as the present epoch when the linear approximation is valid, 
the solution for the dilaton in the $\sigma$-model frame, 
as derived from the equations (\ref{liouvilleeq}), takes the form:
\begin{equation}
\label{dilaton} 
\Phi (t) =-{\rm ln}\left[\frac{\alpha A}{F_1}{\rm cosh}(F_1t)\right],
\end{equation}
where $F_1$ is a positive constant, $\alpha$ is a numerical constant of 
order one, and 
\begin{equation}
\label{defA2}
A = \frac{C_5 e^{s_{01}}}{\sqrt{2}V_6}~, 
\end{equation}
where $s_{01}$ is the equilibrium
value of the modulus field $\sigma_1$ associated with the large bulk 
dimension, and $C_5$ is the corresponding flux of the five-form field.
Notice that $A$ is independent of this large bulk dimension.

For very large times $F_1 t \gg 1$ (in string units),
one therefore approaches a 
linear solution for the dilaton:
$\Phi \sim {\rm const} -F_1 t$.
From (\ref{dilaton}), (\ref{defstringcoupl}) and (\ref{planckstring}),   
we then see that the asymptotic weakness of 
gravity in this Universe~\cite{dgmpp} is due to the smallness of 
the internal space $V_6$ as compared with the flux $C_5$ of the 
five-form field.
The constant $F_1$ is related to the central-charge deficit 
of the underlying the non-conformal $\sigma$-model by~\cite{dgmpp}:
\begin{equation}
\label{ccd}
Q = q_0 + \frac{q_0}{F_1}(F_1 + \frac{d\Phi}{dt}),
\end{equation}
where $q_0$ is a constant, the parenthesis vanishes asymptotically, and 
the numerical solution of
(\ref{liouvilleeq}) studied in \cite{dgmpp}) requires that $q_0/F_1 = (1
+ \sqrt{17})/2 \simeq 2.53$.  For this behaviour of $\Phi$, the
central-charge deficit (\ref{ccd}) tends to a constant value $q_0$. In
this way, $F_1$ is related to the asymptotic constant value of the
central-charge deficit, up to an irrelevant proportionality factor of
order one, in agreement with the conformal model of~\cite{aben}, to which
this model asymptotes. This value should be, for consistency of the
underlying string theory~\cite{aben}, some discrete value for which the 
factorization property (unitarity) of the
string scattering amplitudes is valid. Notice that this asymptotic string
theory, with a constant (time-independent) central-charge deficit, $q_0^2
\propto c^*-25 $ (or $c^*-9$ for superstring)  is considered as an {\it
equilibrium} situation, and an $S$ matrix can be defined for specific
(discrete) values of the central charge $c^*$. The standard critical
(super)string corresponds to a central charge $c^*=25$ ($c^* = 9$), but in
our case $c^*$ differs from that critical value.

Defining the Einstein-frame time $t_E$ through (\ref{einsttime}),
we obtain in the case (\ref{dilaton}) 
\begin{equation}
t_E=\frac{\alpha A}{F_1^2}{\rm sinh}(F_1 t).
\end{equation}
In terms of the Einstein-frame time,
one obtains a logarithmic time-dependence~\cite{aben} for the dilaton
\begin{equation} 
\Phi _E = {\rm const} -{\rm ln}(\gamma t_E)~,
\label{einsteindil} 
\end{equation}
where
\begin{equation}\label{defA} 
\gamma \equiv \frac{F_1^2}{\alpha A}~. 
\end{equation}
For large $t_E$, e.g., present or later cosmological time values, 
one has
\begin{equation} 
a_E(t_E) \simeq \frac{F_1}{\gamma}\sqrt{1 + \gamma^2 t_E^2}.
\end{equation}
At very large (future) times, $a(t_E)$ scales linearly with the
Einstein-frame cosmological time $t_E$~\cite{dgmpp}, and hence the cosmic 
horizon expands logarithmically. From a field-theory viewpoint, this would 
allow for a
proper definition of asymptotic states and thus a scattering matrix.  As
we mentioned briefly above, however, from a stringy point of view, there
are restrictions on the asymptotic values of the central-charge deficit
$q_0$, and it is only a discrete spectrum of values of $q_0$ which allow
for a full stringy S-matrix to be defined, respecting modular
invariance~\cite{aben}. 

Asymptotically in time, therefore, the Universe
relaxes to its ground-state equilibrium situation and the non-criticality
of the string, caused by the initial quantum fluctuation or other initial
condition, disappears, giving way to a critical (equilibrium) string
Universe with a Minkowski metric and a linear-dilaton background.  These
are the generic features of the models we consider here, which can include
strings with target-space supersymmetry as well as the explicit bosonic
Type-0 string considered here for simplicity.

The Hubble parameter of such a Universe becomes for large $t_E$ 
\begin{equation}
\label{hubble2} 
H(t_E) \simeq \frac{\gamma^2 t_E}{1 + \gamma^2 t_E^2}~.
\end{equation}
On the other hand, the Einstein-frame effective four-dimensional 
`vacuum energy density', which is determined by the running 
central-charge deficit $Q^2$ after
compactifying to four dimensions the ten-dimensional
expression $\int d^{10}x \sqrt{-g}e^{-2\Phi}Q^2(t_E)$, 
is~\cite{dgmpp}:
\begin{equation} 
\Lambda_E (t_E) = e^{2\Phi - \sigma_1 - 5\sigma_2}Q^2(t_E) 
\simeq \frac{q_0^2 \gamma^2}{F_1^2 ( 1 + \gamma^2 t_E^2)},
\label{cosmoconst} 
\end{equation}
where, for large $t_E$, $Q$ is given in (\ref{ccd}), 
and approaches its equilibrium value $q_0$. Thus, 
we see explicitly how the dark energy 
density relaxes to zero for $t_E \to \infty$.

Finally, and most importantly for our purposes here, 
the deceleration parameter in the same regime of $t_E$ becomes:
\begin{equation} 
q(t_E) = -\frac{(d^2a_E/dt_E^2)~a_E}{({da_E/dt_E})^2} 
\simeq -\frac{1}{\gamma^2 t_E^2}.
\label{decel4}
\end{equation}
The key point about this expression is that, as is clear from 
(\ref{einsteindil}) and (\ref{defstringcoupl}),
up to irrelevant proportionality constant factors which by 
conventional normalization are set to unity, it can be identified with the 
square of the string coupling:
\begin{equation}
q(t_E) = - {\rm exp}[2(\Phi - {\rm const})] = - g_s^2.
\label{important}
\end{equation}
{\it This is our central 
result.}

To guarantee consistency of perturbation theory, one must have $g_s < 1$,
which can be achieved in our approach 
if one defines the  
present era by the time regime
\begin{equation}
\gamma \sim t_E^{-1} 
\label{condition}
\end{equation} 
in the Einstein frame.
This is compatible with large enough times $t_E$ (in string units) 
for 
\begin{equation} 
|C_5|e^{-5s_{02}} \gg 1~, 
\label{largetwe}
\end{equation} 
as becomes clear from (\ref{defA2}) and (\ref{defA}). 
This condition can be guaranteed
either for small radii of the five extra dimensions 
or by a large value of the flux $|C_5|$ of the five-form 
of the Type-$0$ string. We recall that the relatively large extra 
dimension, $s_{01}$, which extends in the direction of the flux, 
decouples from this condition. Therefore, effective five-dimensional models 
with a large uncompactified fifth dimension may be constructed 
consistently with the condition (\ref{condition}).  

We next turn to the equation of state in such a Universe.
As discussed in~\cite{dgmpp}, this model resembles quintessence,
with the dilaton playing the r\^ole of the quintessence
field. Hence the equation of state
for our Type-$0$ string Universe reads~\cite{carroll}:
\begin{equation}\label{eqnstate} 
          w_\Phi = \frac{p_\Phi}{\rho_\Phi}=\frac{\frac{1}{2}({\dot \Phi})^2 - V(\Phi)}
{\frac{1}{2}({\dot \Phi})^2 + V(\Phi)},
\end{equation}
where $p_\Phi$ is the pressure and $\rho_\Phi$ is the energy density, 
and $V(\Phi)$ is the effective potential for the dilaton, which in our case
is provided by the central-charge deficit term. 
Here the dots denote Einstein-frame differentiation. 
In the Einstein frame, the 
potential $V(\Phi)$ is given by $\Lambda _E $ in (\ref{cosmoconst}).
In the limit $Q \to q_0$, which we assume
characterizes the present era
to a good approximation, $V(\Phi)$ is of order                             
$(q_0^2/2F_1^2)t_E^{-2}$, where we recall 
that $q_0/F_1$ is of order one, as discussed above.
The exact normalization of the dilaton field in the Einstein frame is 
$\Phi _E = {\rm const} -{\rm ln}(\gamma t_E) $.
We then obtain for the present era: 
\begin{equation}
\label{dilpotkin}  
\frac{1}{2}{\dot \Phi}^2 \sim \frac{1}{2t_E^2}, \qquad V(\Phi) 
\sim \frac{6.56}{2}\frac{1}{t_E^2},
\end{equation} 
where the numerical factor is a consequence of the numerical result 
of~\cite{dgmpp}. This implies an equation of state (\ref{eqnstate}): 
\begin{equation} 
w_\Phi (t_E \gg 1) \simeq -0.74 
\label{eqnstatedil}
\end{equation}
for (large) times $t_E$ in string units corresponding to the present era
(\ref{condition}).
Correspondingly, we have a cosmic deceleration parameter
\begin{equation} 
q = \frac{1}{2}(1 + 3w_\Phi) = - 0.61.
\label{valueq}
\end{equation}
This fixes the string coupling to perturbative values, 
consistent with naive scenarios for grand unification.

So far the model did not include ordinary matter, as only fields from
the gravitational string multiplet have been included. The
inclusion of ordinary matter is not expected to change
qualitatively the result. We conjecture that the 
fundamental relation (\ref{important}) will continue to hold,
the only difference being that probably the inclusion of 
ordinary matter will tend to reduce the string acceleration,
due to the fact that matter is subject to attractive gravity,
and resists the acceleration of the Universe. 
In such a case, one has
\begin{equation}
q = \frac{1}{2}\Omega_M - \Omega_\Lambda~,
\label{mattercosmo}
\end{equation}
where $\Omega_M (\Omega_\Lambda)$ denote the matter (vacuum) energy 
densities, normalized to the critical energy density 
of a spatially flat Universe.

There is a remarkable coincidence in numbers for this non-supersymmetric
Type-0 string Universe with the astrophysical observations, which yield
also a $q$ close to this value, since the ordinary matter content of the
universe (normalized with respect to the energy density of a flat
Universe) is $\Omega_{\rm ordinary~matter} \simeq 0.04$ and the dark
matter content is estimated to be $\Omega_{DM} = 0.23$, while the dark
energy content is $\Omega_\Lambda \simeq 0.73$. This yields $q = -0.55$,
which is only a few per cent away from (\ref{valueq}). Conversely, if one
used naively in the expression (\ref{mattercosmo})  the value
(\ref{valueq}) for $q$, obtained in our case where ordinary matter was
ignored, one would find $\Omega_\Lambda \simeq 0.74$, indicating that the
contribution of the dilaton field to the cosmic acceleration is the
dominant one.
  
If the relation (\ref{important})  were to hold also upon the inclusion of
matter, even in a realistic case with (broken) supersymmetry, one would
arrive at a value of the string coupling, $g_s^2 \simeq 0.55$, which would
be quite consistent with the unification prediction of the minimal
supersymmetric extension of the Standard Model at scales $\sim 10^{16}$
GeV. The only requirement for the asymptotic condition (\ref{important})
to hold is that the underlying stringy $\sigma$ model theory is
non-critical and asymptotes for large times to the linear-dilaton
conformal field theory of~\cite{aben}. It should be understood, though,
that the precise relation of the four-dimensional gauge coupling with the
ten-dimensional string coupling depends on the details of
compactification, which we did not discuss in this work.

We close this discussion by stressing once more the importance of
non-criticality in order to arrive at (\ref{important}). In critical
strings, which usually assume the absence of a four-dimensional dilaton,
such a relation cannot be obtained, and the string coupling is not
directly measurable. The logarithmic variation of the dilaton field with
the cosmic time at late times implies a slow variation of the string
coupling (\ref{important}), $|{\dot g_s}/g_s| = 1/t_E \sim 10^{-60}$ for 
the
present era, and hence a correspondingly slow variation of the gauge
couplings.

From a physical point of view, the use of critical strings to describe the
evolution of our Universe seems desirable, whilst non-critical strings may
be associated with non-equilibrium situations, as undoubtedly occur in the
early Universe. The space of non-critical string theories is much larger
that of critical strings. Therefore, it is remarkable that the departure
from criticality has the potential to enhance the predictability of
string theory to such a point that the string coupling may become
accessible to experiment. A similar situation arises in a non-critical
string approach to inflation, in the scenario where the Big Bang is
identified with the collision~\cite{ekpyrotic} of two
D-branes~\cite{brany}. In such a scenario, astrophysical observations may
place important bounds on the recoil velocity of the brane worlds after
the collision, and lead to an estimate of the separation of the branes at
the end of the inflationary period.

The approach of identifying target time in such a framework with a
world-sheet renormaliza- tion-group scale, the Liouville mode~\cite{emn},
provides a novel way of selecting the ground state of the string theory,
which may not necessarily be associated with minimization of energy, but
could be a matter of cosmic `chance'. 
The initial state of our cosmos may correspond to a certain 
`random' Gaussian
fixed point in the space of string theories, which is then perturbed in
the Big Bang by some `random' relevant (in a world-sheet sense) 
deformation, making
the theory non-critical and taking it out of equilibrium from a target
space-time viewpoint. The theory then flows, following a well-defined
renormalization-group trajectory, and asymptotes to the specific ground
state corresponding to the infrared fixed point of this perturbed
world-sheet $\sigma$-model theory. This approach allows for many parallel
universes to be implemented of course, and our world would be just one of
these. Each Universe may flow between a different pair of fixed points, as
it may be perturbed by different operators.  It seems to us that this
scenario is much more attractive (no pun intended) and specific than the
static `landscape' scenario~\cite{Land}, which is currently 
advocated as an attempt to
parametrize our ignorance of the true structure of the string/M theory
vacuum and its specification.

\section*{Acknowledgements}

N.E.M. wishes to thank Juan Fuster and IFIC-University of Valencia 
(Spain) for their interest and support, 
and P. Sodano and INFN-Sezione di Perugia (Italy)
for their
hospitality and support during the final stages of this work.
The work of D.V.N. is supported by D.O.E. grant DE-FG03-95-ER-40917.

\end{document}